# Ultrafast field-driven monochromatic photoemission from carbon nanotubes

Chi Li[1†], Xu Zhou[2†], Feng Zhai[3], Zhenjun Li[1], Fengrui Yao[2], Ruixi Qiao[2], Ke Chen[1], Matthew T. Cole[1], Dapeng Yu[2], Zhipei Sun[4*], Kaihui Liu[2*] Qing Dai[1*]

[1]Nanophotonics Research Division, CAS Center for Excellence in Nanoscience, National Center for Nanoscience and Technology, Beijing 100190, China.

[2]School of Physics, Academy for Advanced Interdisciplinary Studies, Collaborative Innovation Center of Quantum Matter, Peking University, Beijing 100871, China.

[3]Department of Physics, Zhejiang Normal University, Jinhua 321004, China.

[4]Department of Electronics and Nanoengineering, Aalto University, Tietotie 3, FI-02150 Espoo, Finland.

*Correspondence: daiq@nanoctr.cn (Q.D.), khliu@pku.edu.cn (K.L.), zhipei.sun@aalto.fi (Z.S.).

†These authors contributed equally to this work

**Ultrafast electron pulses, combined with laser-pump and electron-probe technologies, allow for various forms of ultrafast microscopy and spectroscopy to elucidate otherwise challenging to observe physical and chemical transitions[1,2]. However, the pursuit of simultaneous ultimate spatial and temporal resolution has been largely subdued by the low monochromaticity of the electron pulses and their poor phase synchronization to the optical excitation pulses[3]. State-of-the-art photon-driven sources have good monochromaticity but poor phase synchronization. In contrast, field-driven photoemission has much higher light phase synchronization, due to the intrinsic sub-cycle emission dynamics, but poor monochromaticity[4,5]. Such sources suffer from larger electron energy spreads (3 - 100 eV) attributed to the relatively low field enhancement of the conventional metal tips which necessitates long pump wavelengths (> 800 nm) in order to gain sufficient ponderomotive potential to access the field-driven regime[4]. In this work, field-driven photoemission from ~1 nm radius carbon nanotubes excited by a femtosecond laser at a short wavelength of 410 nm has been realized. The energy spread of field-driven electrons is effectively compressed to 0.25 eV outperforming all conventional ultrafast electron sources. Our new nanotube-based ultrafast electron source opens exciting prospects for attosecond imaging and emerging light-wave electronics[6].**

Ultrafast electron pulses, pumped by femtosecond lasers, allow for the unprecedented study of various ultra-fast phenomena with high spatial resolution, such as sub-particle ultra-fast spectral imaging[7], real-time protein-protein interactions[8], and electromagnetic waveform microscopy[9]. Nevertheless, it remains challenging to simultaneously extend the spatial and temporal resolution of incumbent electron sources, which are mainly determined by the electron beam energy spread, which determines the beam spatial resolution, and high phase synchronization between the probe electron pulses and the pump light/electron pulses, which determines the temporal resolution[3,10]. It has proven difficult to date to simultaneously realize these two properties in incumbent electron source system, via both photon-driven (quantum) or field-driven (classic) approaches.

Photon-driven photoemission sources are widely applied in state-of-the-art time-resolved electron microscopies and spectroscopies, as their highly monochromatic emission (~0.7 eV of energy spread[6]) enables high spatial resolution (sub-nanometer) through photon energy - work function matching[10]. However, the temporal resolution (sub-picosecond) of the microscopies and spectroscopies is limited as the photon-driven emission is not well synchronized to the optical phase[1]. Although great efforts have been devoted to improve the temporal resolution, through phase modulation in tailored near-field nano-cavities following the photoemission process[6,11] for example, the ultimate performance of the modulated electron beam still largely limited by the initial beam emission dynamics.

In contrast, field-driven photoemission typically occurs through sub-cycle durations, which naturally achieves electron pulses with high optical phase synchronization[4,5]. Indeed, this field-driven photoemission has been previously achieved from sharp metal tips under infrared excitation (wavelength > 800 nm)[12,13]. Unfortunately, the electron energy spread of such sources is extremely high (3 - 100 eV[4,5,13]) making them ill-suited for many emerging samples of interest. In this regime, the narrow electron energy spread strongly depends on the high field localization and short excitation wavelength. Ultra-sharp tips are essential to reduce the electron energy spread. Sharper tips give rise to higher field localization, and facilitate access to the field-driven photoemission regime at shorter wavelengths due to notably stronger field enhancement. However, this is difficult to achieve from traditional metal tips due to difficulties associated with their manufacturing coupled to a need for a sufficiently robust material with a high damage threshold required to survive the high optical fields. The ultra-small tip radius (~ 1nm), strong field enhancement and impressive structural stability (melting point >2000 K in vacuum), of carbon nanotubes (CNTs), have traditionally afforded highly coherent field electron emission capable of low energy spreads, small virtual source size and high brightness[14-19]. As a static electron emmision source, CNTs have been shown to outperform metal tips across almost all qunatifiable metrics[14].

In this work, we demonstrate an ultrafast carbon nanotube-based, sub-nanometer electron source. The extremely sharp tips, coupled to their structural stability allow for extremely high optical field localization, thereby enabling access to the field-driven photoemission regime at unprecedentedly short wavelengths as low as 410 nm, compared to 800-3000 nm as conventionally used for metallic tips[4,12]. As a result, a much narrower energy spread (~0.25 eV) is obtained, an entire order of magnitude improved over previously demonstrated field-driven electron sources[4,5]. The photoemission process is schematically illustrated in Fig. 1A. In our experiments, CNTs were grown by chemical vapor deposition (see Methods for details) and had a narrow normally distributed nominal tube radius between 0.5 - 1 nm (Fig. 1B, C, see Methods for further details of sample preparation and characterizations).

In the experiments, femtosecond pulses were focused onto the CNT tips under ultrahigh vacuum, as illustrated in Fig. 1D (see Methods for details). Photoemission from the CNTs was found to depend strongly on the angle of linear polarization of the incident laser, as shown in Fig. 2A. The polarization dependence exhibits a $\cos^6\theta$ behavior, where $\theta$ is the angle between the tip shank and the polarization of the input optical source. Such optically driven nonlinear photoemission suggested as the observed translational symmetry only allows excitation by the field component vertical to the emitting surface[15,16]. This cosinsoidal curve also largely excludes any possibility of thermally induced field emission, which has been shown elsewhere to display a sinusoidal-like dependence on the polarization angle[17].

Photoemission currents (I) were recorded as a function of incident laser power (P), as shown in Fig. 2B. The emission current at low pump power was approximately proportional to the fourth power of laser power, suggesting above threshold multiphoton photoemission[18]. The measured work function (Φ) of the CNT (~ 4.4 eV, see Methods and Extended Data Fig. 1 for details) requires only two photons for photoemission. The enhanced emission nonlinearity indicates that electrons are sourced almost exclusively from the sharp CNT tips[12]. A downward deviation from the fourth power law to third power law was found as the pump power increases, which presents transition from above threshold multiphoton photoemission to classical field-driven photoemission[12,20]. The third power law of the I-P curve is also consistent with the $\cos^6\theta$ dependency of the polarization angle in Fig. 2A, as the optical field amplification adopts a square-law relation to the laser power. The Fowler-Nordheim (FN) plot[21] derived from the above current-field curve demonstrates a high degree of linearity, as shown in Fig. 2C, which further confirms that the dominant emission has field-driven behavior, according to cycle-averaged FN model. The field enhancement factor (β) can be calculated from the slope (*S*) of a linear fit[22], where $\beta = -B\Phi^{\frac{3}{2}}/S$. Here B is the constant 6.83×10$^9$ V eV$^{-3/2}$ V m$^{-1}$. The calculated field enhancement factor was ~27 (approaching the value estimated from our finite element analysis, see Methods, Electromagnetic simulations), which is approximately twice as large as those of conventional metal nanotips[4,5,13]. This explains why the visible field-driven wavelength operation is feasible with our CNT emitter, whilst near-infrared excitation is required for metallic tips.

For optical frequency field emission, the Keldysh parameter (γ) has been used to estimate the magnitude of the optical field required to support quasistatic electron tunneling, where γ < 2 describes tunneling emission, with the transition to tunneling behavior usually occurring in the range 1 < γ < 2.[18] In the present case, for a 3 mW 410nm incident laser, the calculated γ was ~1.85. Thus, both the cycle-averaged FN fitting and calculated Keldysh γ, support access to ultrafast photoemission into the field-driven tunneling regime for the present carbon nanotube system. This provides an effective and convenient means of realizing narrow electron energy spreads. We stress that this regime was previously unattainable at visible wavelengths due to the relatively low field enhancement and damage threshold of conventional metal tips.

Beam characterization was conducted using a retarding field method achieved by scanning the anode bias voltage under varying laser powers (I-V measurements, Fig. 3A). Three electron dynamics stages (marked as I, II, III in Fig. 3A) were involved in the measurements. In stage I (the left-most region), all the emitted electrons exist in a fully retarding field (negative anode bias), such that no electrons can reach the anode as their kinetic energy cannot overcome the retarding potential. As the anode bias increases, the energetic electrons increasingly penetrate through the reduced retarding potential, and ultimately reach the anode (stage II, highlighted in pink in Fig. 3A). In this regime, electrons with different kinetic energy require different collection potentials, such that the width of the potential range in this stage reflects the kinetic energy spread of the electrons, which behave increasing with the laser power increasing, in good agreement with theoretical descriptions reported elsewhere[4]. Due to the anisotropic local field around the tip alongside possible Coulomb repulsion, the emitted electrons diverge to form a cone beam[12], which will be continuously focused and collected under the action of the increasing anode bias. This leads to a reduced rate of increase in the current, but still depends nonetheless on the laser power, as observed in stage III (the right most region). The slope increases with the laser power, indicating that high optical fields also broaden the electron emission angle. We note a constant current at the right region at an incident laser power of 3 mw suggesting that the generated electron beam has a very small emission angle.

The energy distribution of the photo-generated electron beam can be determined directly from the differential spectrum of the above I-V curves[23], as shown in Fig. 3B. The peaks are derived from stage II, of which the width reflects the kinetic energy spread and demonstrates a clear dependency on the laser power. The right shoulders of the peaks are derived from stage III, of which the height reflects the beam divergence. As noted above, when the pump power was 3 mW, a very narrow energy spread down to 0.25 eV (full width at half maximum, FWHM, Gaussian fitting) was observed, which is more than one order of magnitude smaller than that of previously reported field-driven photoemission from metal tips (3-100 eV), and at least two times smaller than previously reported photon-driven sources[6].

To elucidate this emission behavior, we compute the kinetic energy spectrum using experimentally derived parameters in an extended two-step Simpleman model[4]. The simulation includes a simplified FN tunneling model (first step - electron tunneling), and the interaction of the electrons with a strongly localized field near the tip (second step - electron propagation) (see Methods for details, Simulation of kinetic energy spectrum). Fig. 3C shows the computed contour plot of individual kinetic energy spectrum as a function of laser power. The spectrum features are strongly modulated with the laser power: the FWHM (the range between the upper and lower dotted lines), refer to the electron energy spread and how this increases as a function of laser power, which is consistent with the experimental results observed in Fig. 3B. At 3 mW pump (purple dash line), our simulations reveal a kinetic energy spread of ~ 0.24 eV, which is in good agreement with the experimental data (0.25 eV). The narrow energy spread and the small divergence of the electron pulses greatly benefit beam line collimation and compression for functionally enhanced microscopy and spectroscopy. We note a broadening of the experimental energy spread relative to our simulations with increasing laser power (inset of Fig. 3C). This may, in part, be attributed to enhanced tunneling probability near Fermi level energies when exposed to high optical fields, which will contribute to the final measured energy spread[25].

By calculating γ as a function of pump wavelength at a fixed surface intensity (56 GW/cm$^2$) for CNTs with *β*=27 and metal tips with *β*=10 (from references[4,5,25]), we realized that the higher capability of CNTs to access into field-driven photoemission under a much shorter pump wavelength than that of metal tips (Fig. 3D). This suggests CNTs are an excellent platform for realizing broadband optical frequency-based electron emitters. In addition, electron pulses, in the single/few electron mode[3,26], operating continuously for more than 50 hours (see Extended Data Fig. 3) at a high repetition rate of 80 MHz without obvious degradation, highlights the potential of the present CNT devices to operate as high brightness, long lifetime ultrafast electron sources.

Ultrafast processes, such as electronic transitions at the atomic scale, can evolve on time scales of a few femtoseconds and below. Advancing ultrafast imaging into this regime requires electron pulses of not only attosecond duration, but also high optical phase synchronization. Pulses generated from a traditional photon-driven photocathode cannot be shorter than the optical femtosecond pulses used for photoemission, and have a poor optical phase synchronization. In contrast, field-driven photocathodes provide a much shorter pulse duration due to the sub-cycle emission process, which lead to near ideal optical phase synchronization.

Here we have demonstrated field-driven photoemission from CNT emitters capable of generating extremely low energy spread (~0.25 eV), which outperforms the current state-of-the-art field-driven electron sources by at least an order of magnitude. The high optical field enhancement (~27) in the engineered CNTs allows, for the first time, access to field-driven photoemission at unprecedently short wavelengths (410 nm) - an issue that continues to plague metallic sources - which provides much improved beam coherence. Combining the unique geometrical and metrological properties of CNTs allows for simplified beam line collimation and compression system design in ultrafast microscopies and spectroscopies. The present findings suggest nanoengineered CNTs are a likely rich source for the realization of ultrafast electron guns, with the present findings going someway in contributing to attosecond resolution electron microscopy and spectroscopy, as well as enabling new findings in ultrafast science and emerging designs for the next-generation of light-wave electronics.

**Acknowledgments** This work was supported by the National Basic Research Program of China (grant no. 2016YFA0202000, 2015CB932400, 2016YFA0300903), the National Natural Science Foundation of China (grant no. 11427808, 51372045, 11474006, 5152220, and 91433102), the International Science and Technology Cooperation Project (no. 2014DFR10780, China), the Academy of Finland (grant no: 276376, 284548, 295777, and 304666), TEKES (OPEC), Nokia foundation, the Oppenheimer Trust, and the European Union's Seventh Framework Programme (grant no: 631610).

**Additional Information** Reprints and permissions information is available at www.nature.com/reprints. The authors declare no competing financial interests. Readers are welcome to comment on the online version of the paper. Correspondence and requests for materials should be addressed to Q.D. (daiq@nanoctr.cn), K.L. (khliu@pku.edu.cn), Z.S. (zhipei.sun@aalto.fi).

**Figures and Legends**

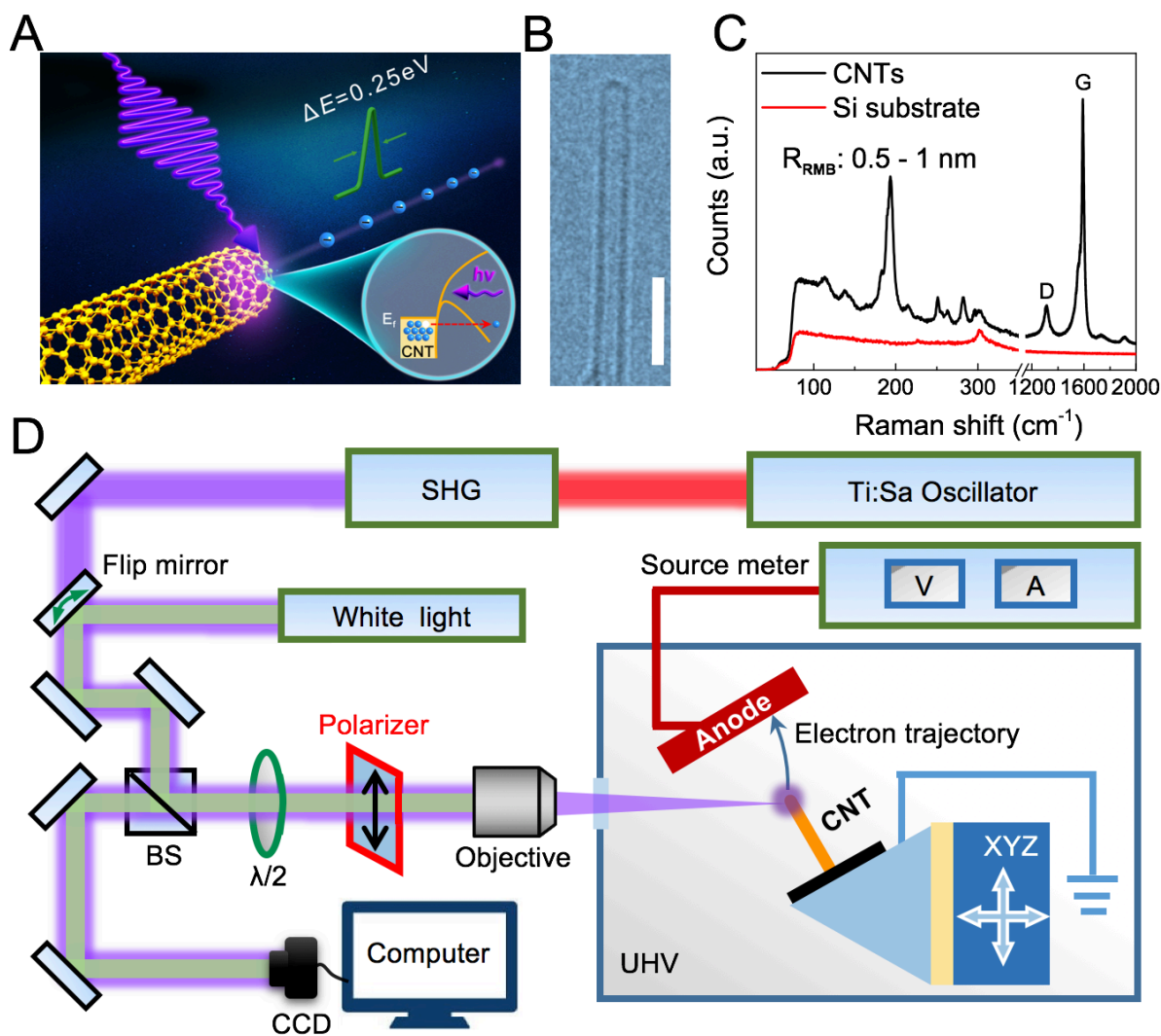

**Figure 1 | Highly coherent CNT-based photoemission source.** (**A**) Emission dynamics. (**B**) High-resolution transmission electron microscopy image of a typical CNT under study. Scale bar: 5 nm. (**C**) Raman spectrum of a typical as-grown CNT samples. (**D**) Experimental setup. (SHG, second harmonic generation. CCD, charge-coupled device. BS, beam splitter. UHV, ultra-high vacuum.)

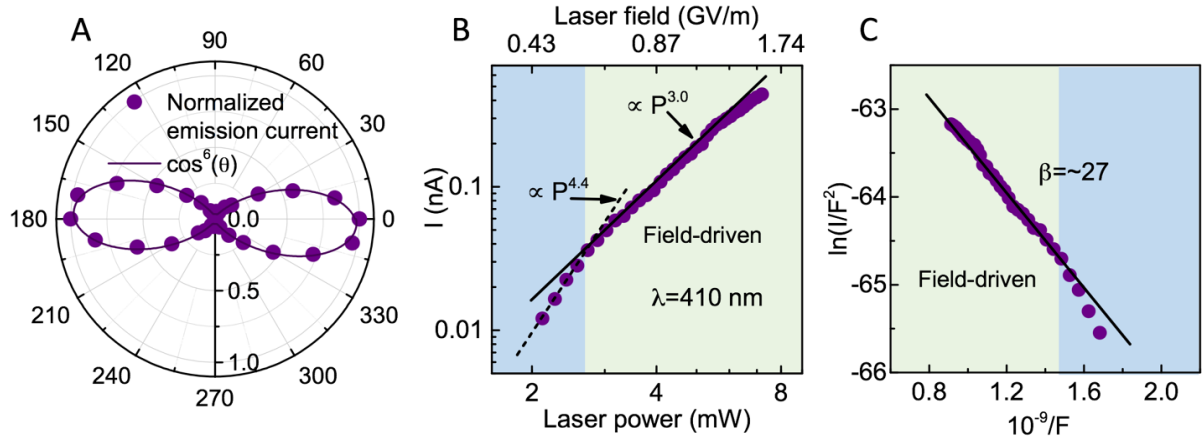

**Figure 2 | Field-driven photoemission at 410 nm pump wavelength.** (**A**) Normalized emission current as a function of the angle of polarization (θ) of the input optical source. Note the emission currents show a $\cos^6(\theta)$ dependence. (**B**) Emission current as a function of laser power (bottom abscissa) and laser field (top abscissa). At low power range, fourth power is noted, while third power law is noted at higher power range. (**C**) FN fitting of the optically driven emission current, with a field enhancement factor (β) of ~27. Green area in (B) and (C) corresponds to the field-driven regime region

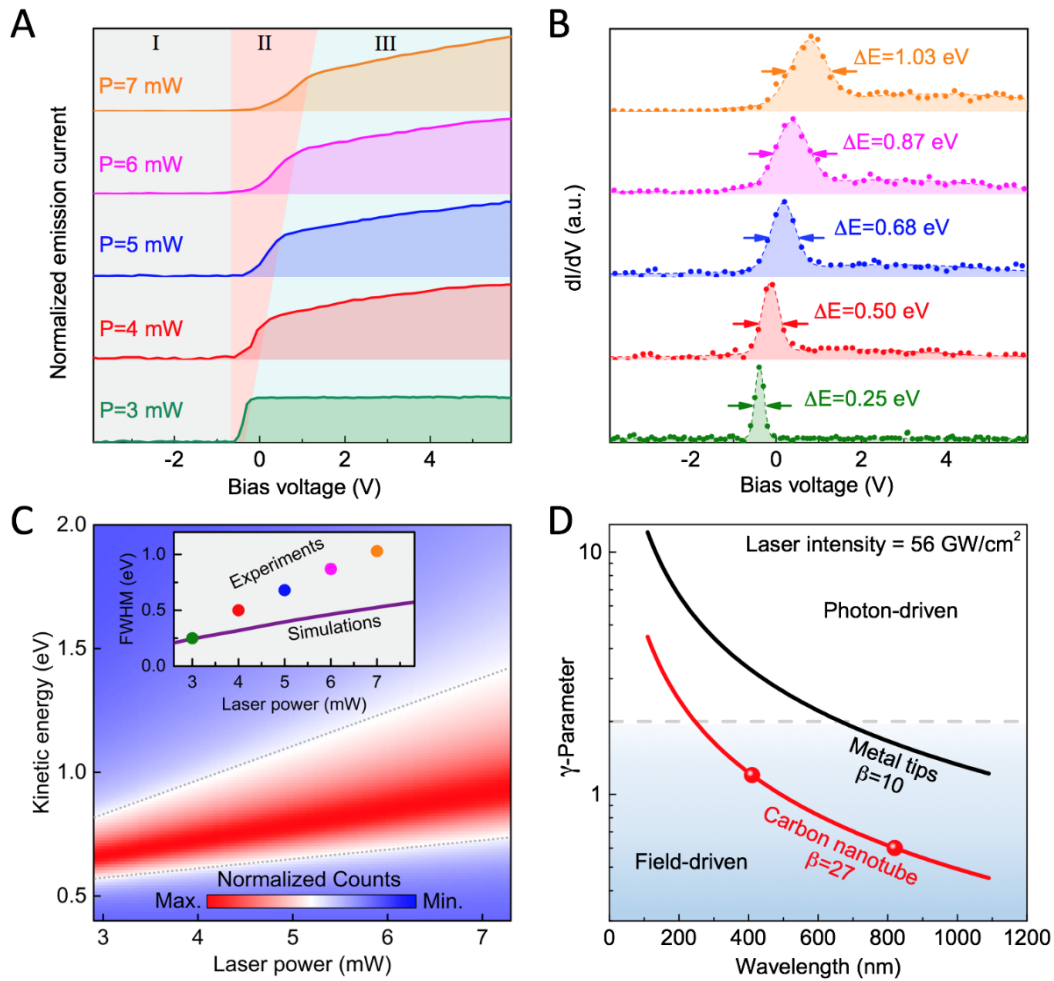

**Figure 3 | Electron beam characterization.** (**A**) Dependence of the normalized emission current on bias voltage (I-V) at various incident powers. Stage I: fully retarding region; stage II: fast collecting region; stage III: beam focusing region. (**B**) dI/dV curves at different power; the width of the peaks (FWHM) indicates the total energy spreads, while the shoulder indicates the beam divergence grade. (**C**) Contour plot representation of the normalized counts as a function of laser power and kinetic energy, according to the extended Simpleman model. The FWHM is marked as the range between upper and lower dotted lines. The inset shows FWHM of the experimental energy spread (colour dots) and simulated kinetic energy spectrum (purple line). (**D**) Calculated Keldysh parameter $\gamma$ as a function of wavelength at given field enhancement factor for carbon nanotube (~ 27) and metal (~ 10) and fixed laser intensity of 56 GW/cm$^2$ (corresponding to 7 mW in present work). The two reds dots correspond to the results of this work.

**Methods**

**Growth of single wall carbon nanotubes.** Vertically aligned single walled CNT arrays were grown on highly doped n-type silicon chip by chemical vapor deposition (CVD). The silicon substrate was first coated with an Al (10 nm) / Fe (1 nm) multilayer catalyst, deposited by sputtering. The substrate was then heated to 900 °C, at $10^{-2}$ mbar. During heating, gaseous ammonia was introduced to etch the surface of the nickel catalyst and stimulate the formation of nano-islands which template the induced nanotube self-assembly process. Acetylene was chosen as the carbon feedstock, and introduced to the deposition chamber once the temperature reached 900 °C. The growth process lasted for 1 minutes resulting in 10 μm tall CNT clusters of defined areal patterns. Following the growth process, the samples were annealed in hydrogen at 1000°C for two hours to remove amorphous carbon deposits along with any other remaining impurities.

**Characterization of single wall carbon nanotubes.** A high resolution transmission electron microscopy (HRTEM, FEI Tecnai F20) image is shown in Fig. 1B. The diameters of the CNTs were assessed from their radial breathing mode frequency ($\omega_{RBM}$=248/d (cm$^{-1}$/nm)) through Raman spectrum[27,28], as shown in Fig. 1C. The pristine SWNTs were dispersed in absolute alcohol via ultrasonication and drop transferred onto 300 nm $SiO_2$/Si substrates. Raman spectra were acquired using a He-Ne laser (632.8 nm) excitation, with data recorded using a confocal micro-Raman spectrometer (HORIBA Jobin Yvon, LabRam HR 800) with 0.35 cm$^{-1}$ resolution using 1800 gr/mm grating. The system had a 1 μm optical probe diameter using a 100 x (N.A.=0.9) Nikon objective.

**Ultraviolet Photoelectron Spectroscopy (UPS) was used for for work function assessment.** CNTs were stored in a vacuum desiccator and exposed only briefly to air at STP prior to loading into an ultra-high vacuum chamber equipped with an angle-resolved electron energy analyzer, fitted with a He discharge lamp for UPS measurements. The system was also equipped with a Mg Kα X-ray source, a heating stage and a sputter-cleaning facility, pumped to a base pressure of $2\times10^{-10}$ Torr. During UPS measurements, the system pressure was $10^{-9}$ Torr following He back filling. A He discharge source (21.22 eV) was used with an energy resolution of ~0.2 eV. The work function of the present CNTs was ~4.4 eV, as shown in Extended Data Figure 1.

**Photoemission experimental setup.** A schematic depiction of the experimental electron emission setup is shown in Fig. 1D. Photoelectron emission from CNT arrays was triggered with 100 fs laser pulses, with a central wavelength of either 820 nm or 410 nm, at an 80 MHz repetition rate from a Ti: Sapphire ultrafast laser (Spectra-Physics, Mai Tai-Series, SHG). A standard Si photodiode power sensor (Thorlabs, Photodiode Power Sensor S120C) was used to measure the laser power. White light and a charge coupled device (CCD) were employed to monitor the sample position and the laser spot profile. The laser was linearly polarized with its polarization angle controlled via a polarizer and a half-wave plate. The laser was normally incident on the CNT tip via front illumination, which was focused to a 1.25 / 2.50 μm (FWHM, 410 / 820 nm) spot at the CNT cluster apex. Although the as-grown clusters contain many nanotubes, the growth kinetics were such that a few individual tubes protruded, repeatedly between growths, from these clusters producing a few isolated nanoscopic apex, which we believe are the main photoemission sites giving the extremely high field enhancement there. These photocathode samples were mounted in a high-vacuum chamber ($10^{-7}$ Torr). The anode was adjacent to the photocathode, some 400 μm distant, using a thick mica insulating spacer. The anode, together with the insulating separator, was placed directly on the surface of the photocathode with the CNT arrays centrally aligned. A Keithley 2400 source measurement unit was used to bias the anode with voltages of up to 50 V, with the anode current measured. Unless otherwise stated, the current measurements presented in the main text are those recorded at the anode. Every current data, collected by source meter, was acquired from an arithmetic average of 100 repeated measurements.

**Electromagnetic simulations.** Electromagnetic simulations (Extended Data Figure 2) were conducted using a Finite Element Method coupled to an electrostatic (es) and transient electromagnetic wave (ewt) module. The CNT, copper electrodes and vacuum were modeled in three dimensions. On the top and bottom surfaces of the vacuum domain, the electric potential and perfect electric conductor boundary conditions are applied in the es and ewt modules in order to provide a dc electric field bias and symmetric magnetic fields, respectively. Scatter and periodic boundary conditions are applied on the left/right and front/back surfaces to simulate the incident Gaussian/TE/TM wave. The electromagnetic properties of the CNT are based on those extracted from experimental data found elsewhere[29]. Near-field distributions were obtained by subtracting the incident and scattered plane waves from the total simulated field employing a time-dependent solver.

**Simulation of kinetic energy spectrum.** A two-step Simpleman model[4] was adopted to simulate the kinetic energy spectrum of the liberated electron population. The first step here is electron tunneling. In the quasi-static approximation, the tunneling probability, at time t, is given by the Fowler-Nordheim equation[30], as;

$$P(t) = \Theta[F_{z=0}(t)] \frac{e|F_{z=0}(t)|^2}{16\pi^2 \hbar \Phi} exp\left(\frac{-4\sqrt{2m}\Phi^{3/2}}{3\hbar|F_{z=0}(t)|}\right)$$

Here, $\Theta(x)$ is the Heaviside step function, $m$ and $-e$ are the rest mass and charge of an electron, $\hbar$ is the reduced Plank constant, and $\Phi$ (= 4.4 eV) is the work function of the CNT. The oscillating electric field $F_{z=0}(t)$ at the tip surface (z=0) induces electron emission only when it points outwards from the surface ($F_{z=0}(t) > 0$). The incident laser pulse has a time dependence, as;

$$E_{z=0}(t) = \frac{F_{z=0}(t)}{-e\alpha} = -E_0 \cos(\omega t + \phi)\exp(-2ln2\times t^2/\tau^2)$$

where $\alpha$ is the geometry-dependent field enhancement factor, $E_0$, $\omega$, $\phi$, and $\tau$ are, respectively, the peak strength, circular frequency, carrier-envelope phase, and pulse duration of the incident laser field.

The second step in the Simpleman model is the ponderomotive acceleration of the emitted electrons in the total electric field $E_{tot}(z,t)$. This field depends on both the distance $z$ to the tip and the time *t*. A simplified expression[4] is adopted for $E_{tot}(z,t)$,

$$E_{tot}(z,t) = E_{z=0}(t)\left[(\alpha-1)\left(\frac{r_0}{r_0+z}\right)^3 + \exp\left(\frac{-2ln2\times z^2}{(2w_{foc})^2}\right)\right] + E_{static}(z)$$

where $r_0$ is the tip radius and $w_{foc}$ is the beam waist. The peak strength and $1/e$ decay length of the total field are $E_{tot}^{max} = \alpha E_0$ and $l_F \approx 0.4 r_0$. For CNT tips, the field enhancement varies with the radius as $\alpha = 24(r_0[nm])^{-0.5}$. The static field $E_{static}(z) = -\frac{dV}{dz}$ accounts for the image potential which was modeled as[31]

$$V(z > 0) = (E_F + \Phi) \exp(-\beta r) / [\frac{4z(E_F + \Phi)}{A} + 1]$$

Here $E_F$ is the Fermi energy, the constant $A = \frac{e^2}{4\pi\varepsilon_0} = 1.44\ eV \cdot nm$, and $\beta$ is the screening constant.

For an electron emitted at time $t_B$, the equation of motion $\ddot{z}(t) = -eE_{tot}(z,t)/m$ is numerically integrated by means of the 4$^{th}$ order Runger-Kutta method, with the initial condition $z(t_B) = \dot{z}(t_B) = 0$. Rescattering events at the tip are treated as perfect elastic collisions, with a reflection probability R (R = 0.1 for present case). From the evaluated single-particle trajectory, we obtain the final kinetic energy $\varepsilon_K(t_B, \phi, r_0)$ which depends on the emission time $t_B$, carrier-envelope phase $\phi$, and tip radius $r_0$.

The kinetic energy spectra $f(\varepsilon)$ of final electrons is obtained by weighting the final kinetic energies with the instantaneous generation probability and averaging over the carrier-envelope phases,

$$f(\varepsilon) \propto \int_0^{2\pi} d\phi \int dt_B\, p(t_B) \frac{\Gamma/\pi}{[\varepsilon - \varepsilon_K(t_B, \phi, r_0)]^2 + \Gamma^2}$$

where $\Gamma$ is the energy resolution.

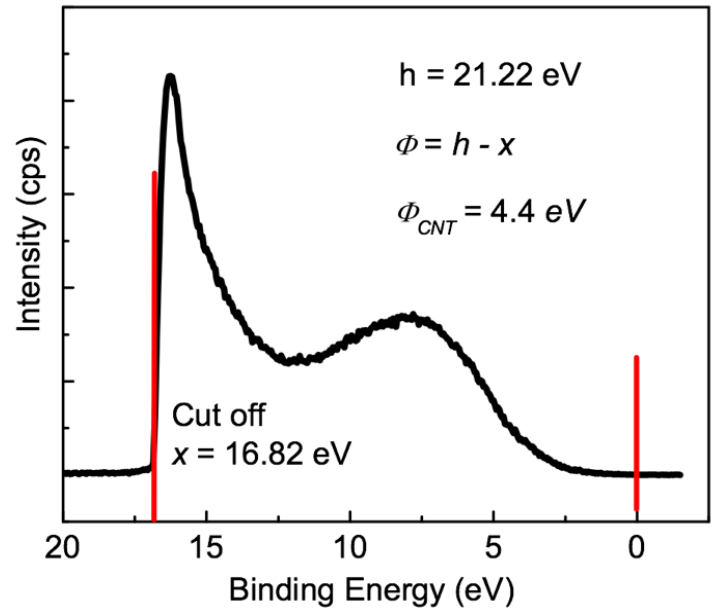

**Extended Data Figure 1 | Work function measurement of CNTs employing UPS.**

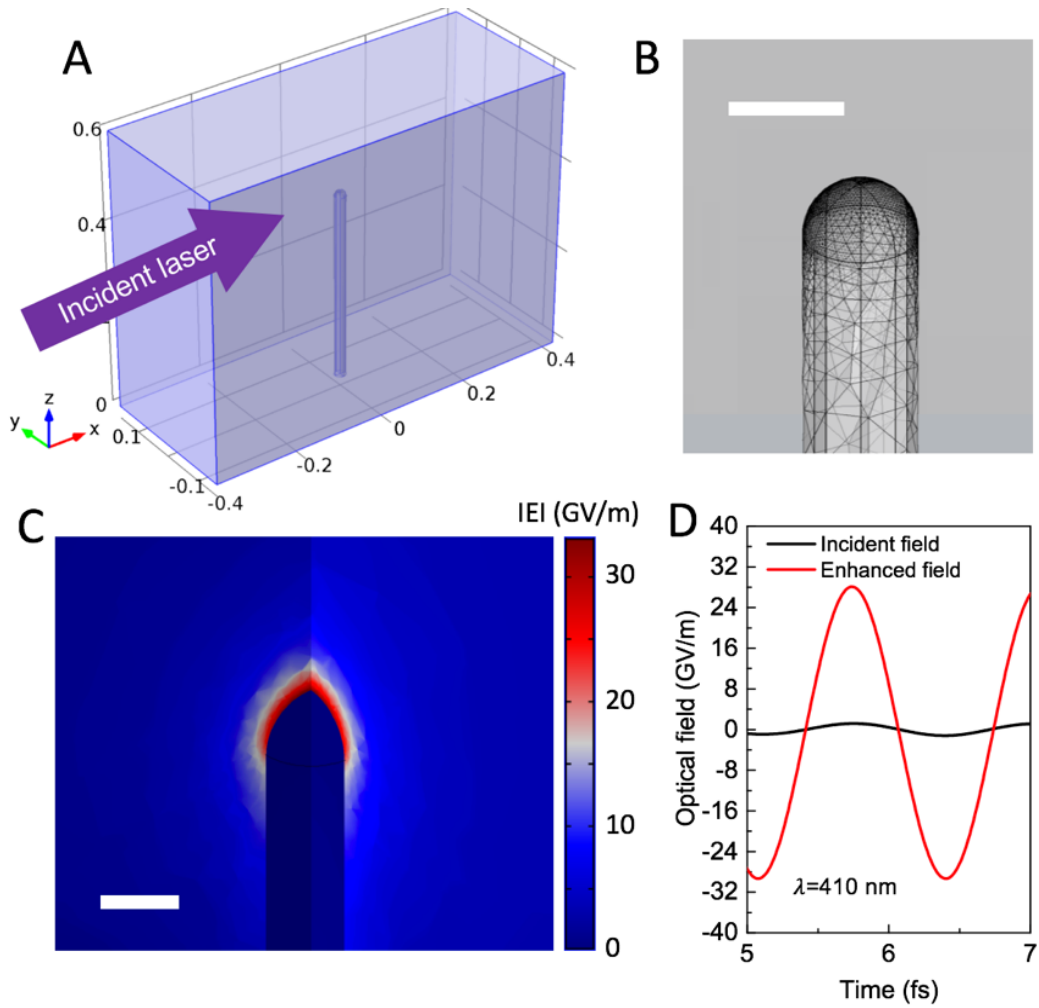

**Extended Data Figure 2 | Electromagnetic simulation. (A)** Simulation mode illustrating the laser incident direction. The CNT is modeled as a passive quasi-metal with length of 100 nm and radius of 1 nm, whose apex is modeled as a semi-sphere (as shown in Fig. 1B). **(B)** Meshing of the CNT apex. To save the simulation time, the apex is specially finely meshed. Scale bar, 2 nm. **(C)** Simulated electric field (modulus) distribution on the apex. Scale bar, 2 nm. **(D)** Time dependent incident field and enhanced field on the apex at 410 nm, which show a field enhancement factor of ~25.

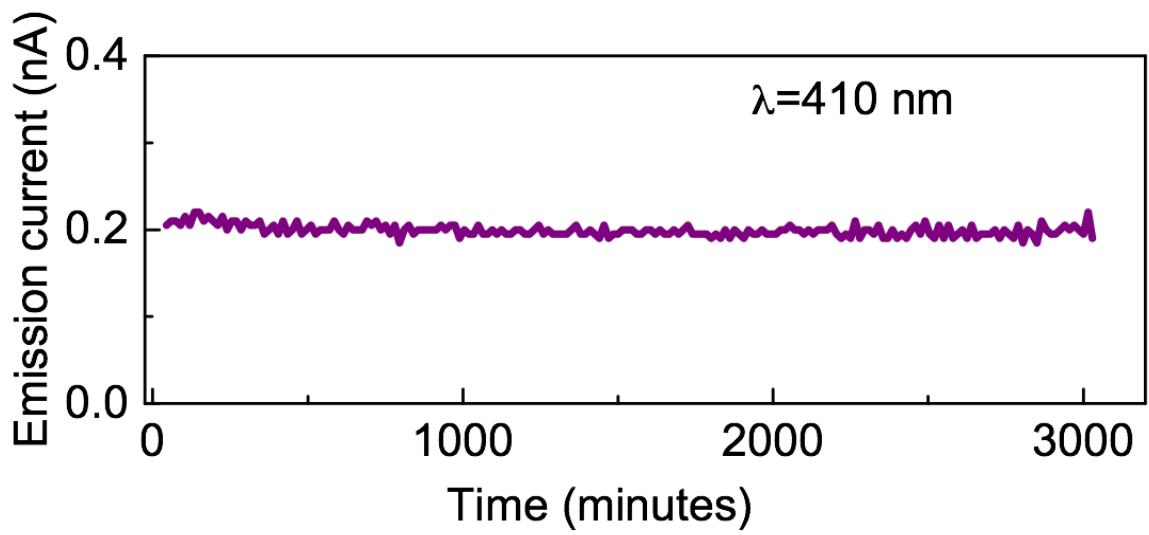

**Extended Data Figure 3 | Photoemission stability test.**